# Transverse energy flow and the "running" behaviour of the instantaneous field distribution of a light beam

**A. Ya. Bekshaev**

*I.I. Mechnikov National University, Dvoryanska 2, Odessa 65082, Ukraine*

*bekshaev@onu.edu.ua*

**Abstract.**
It is known that the orbital angular momentum of a paraxial beam is related to the rotational motion of the instantaneous field pattern representing the electric field optical-frequency oscillations [arXiv:0812.0888; J. Opt. A: Pure Appl. Opt **11**, 094004, (2009)]. Now this conclusion is generalized: any identifiable directed motion of the instantaneous 2D pattern of the field oscillation ("running" behaviour of the instant oscillation pattern) corresponds to the transverse energy flow in the experimentally observable time-averaged field. The transverse orbital flow density can be treated as a natural geometric and kinematic characteristic of this running behaviour.

Usually, paraxial light beams are characterized by the energy or momentum (Poynting vector) distributions averaged over the oscillation period. However, the pattern of the instantaneous field oscillations, which evolve with optical frequency and are normally unobservable, show some remarkable parallels to the immediately observable characteristics of the beam energy transportation. In particular, the orbital angular momentum of a paraxial beam is associated with the rotational motion of the instantaneous field pattern and, moreover, appears to be a direct geometrical and kinematic measure of this rotational motion [1, 2].

In this note we intend to generalize the statements of Refs. [1,2] and to show that any identifiable directed motion of the instantaneous 2D pattern of the field oscillation ("running" behaviour of the instant oscillation pattern) corresponds to the transverse energy flow (or, which is almost the same [3], transverse momentum) of the time-averaged field, and that the transverse energy flow density provides a natural geometric and kinematic characterization of this running behaviour. The idea seems rather simple and transparent; however, to the best of our knowledge, it was never formulated in a clear and explicit form.

Our subject is a scalar monochromatic (wavenumber $k$ and frequency $\omega = ck$) paraxial beam propagating along axis $z$. In the beam cross section, Cartesian $\mathbf{r} = (x, y)$ and polar $r = \sqrt{x^2 + y^2}$, $\phi = \arctan(y/x)$ frames are introduced. The electric field of this beam can be expressed as

$$\mathcal{E}(\mathbf{r},z,t) = u(\mathbf{r},z)\exp[i(kz-\omega t)] \quad (1)$$

where $u(\mathbf{r},z)$ is the slowly varying complex amplitude [3]. By introducing the real modulus $A(\mathbf{r})$ and phase $k\varphi(\mathbf{r})$ of the complex amplitude in accordance with equation

$$u(\mathbf{r}) = A(\mathbf{r})\exp[ik\varphi(\mathbf{r})] \quad (2)$$

the transverse orbital flow density (OFD) [3] of this beam can be expressed in the form [3]

$$\mathbf{S}_{O\perp}(\mathbf{r}) = \frac{c}{8\pi} A^2(\mathbf{r})\nabla\varphi(\mathbf{r}). \quad (3)$$

where $\nabla = \mathbf{e}_x \frac{\partial}{\partial x} + \mathbf{e}_y \frac{\partial}{\partial y}$ is the transverse gradient, $\mathbf{e}_x$ and $\mathbf{e}_x$ are the unit vectors of the transverse coordinate axes. Note that the spin flow [3] in scalar beams is absent and Eq. (3) describes also the total transverse energy flow in the beam cross section.

Equation (1) provides complex representation of a harmonically oscillating field; the true instant field distribution at any time moment is determined by the expression

$$E(\mathbf{r},t) = \text{Re}[\mathcal{E}(\mathbf{r},t)] = A(\mathbf{r})\cos[k\varphi(\mathbf{r}) - \omega t]. \quad (4)$$

This equation describes the instantaneous beam pattern discussed in Refs. [1,2] and in this paper. The constant-level points $\mathbf{r}_c$ of the cosine argument in (4) obey the equation

$$\varphi(\mathbf{r}_c) = ct \quad (5)$$

which, in general, express the running behaviour of the instantaneous oscillation pattern (animated illustrations can be found in [1,2]). For example, in case of circular Laguerre-Gaussian modes [4], in the waist cross section, $A(\mathbf{r}) \equiv A_l(r)$, $k\varphi(\mathbf{r}) = l\phi$ where $l$ is the integer mode index. Then

$$E(\mathbf{r},t) = A(r)\cos(l\phi - \omega t), \quad (6)$$

whence the rotation of the instant beam pattern around the nominal beam axis $z$ is seen evidently [1,2]. Another and even more simple example is the inclined plane wave propagating in direction specified by small angles $k_x/k$ and $k_y/k = 0$. For it $A(\mathbf{r}) \equiv \text{const}$ and $k\varphi(\mathbf{r}) = k_x x$

$$E(\mathbf{r},t) \propto \cos(k_x x - \omega t) \quad (7)$$

which describes the running wave directed along axis $x$ (in accordance with the plane wave inclination). Of course, these running motions possess no direct mechanical meaning (there are possible superluminal velocities, etc.) but we will show that it is immediately related with the transverse energy flow, i.e. with the genuine mechanical momentum of the beam.

To this purpose we employ the same way of reasoning that was used in Refs. [1,2]. Let us find an analytical criterion that the pattern of the instantaneous field oscillations, generally rather complicated, really shows certain running features. First of all, note that in simple cases of Eqs. (6) ("pure" rotation) and (7) ("pure" translation) the following relations take place:

$$\frac{\partial E(\mathbf{r},t)}{\partial t} = -\frac{\omega}{l}\frac{\partial E(\mathbf{r},t)}{\partial \phi}, \quad \frac{\partial E(\mathbf{r},t)}{\partial t} = -\frac{\omega}{k_x}\frac{\partial E(\mathbf{r},t)}{\partial x} \quad (8)$$

– the time and coordinate derivatives of the instant field are proportional. One may expect that in general case, a certain correlation between the quantities

$$\frac{\partial E(\mathbf{r},t)}{\partial t} \text{ and } \nabla E = \mathbf{e}_x \frac{\partial E(\mathbf{r},t)}{\partial x} + \mathbf{e}_y \frac{\partial E(\mathbf{r},t)}{\partial y}$$

will testify for the running behaviour discussed.

A usual measure of the correlation between two functions is the mean value of their product which, for periodic functions considered here, can be replaced by the temporal mean value over the oscillation period. Thus, the sought correlation coefficient equals to

$$\langle E_{\mathbf{r}} E_t \rangle \equiv \left\langle \nabla E(\mathbf{r},t) \frac{\partial E(\mathbf{r},t)}{\partial t} \right\rangle = \frac{1}{T} \int \nabla E(\mathbf{r},t) \frac{\partial E(\mathbf{r},t)}{\partial t} dt \qquad (9)$$

where $\langle \ldots \rangle$ means averaging, $T = 2\pi/\omega$ is the optical oscillation period. Taking Eq. (4) into account, the latter relation can be transformed to

$$\langle E_{\mathbf{r}} E_t \rangle = -\frac{k\omega}{2} A^2(\mathbf{r}) \nabla \varphi(\mathbf{r})$$

which, in view of (3), immediately gives

$$\langle E_{\mathbf{r}} E_t \rangle = -4\pi k^2 \mathbf{S}_{O\perp}. \qquad (10)$$

Therefore, the correlation coefficient (9) that, basing on geometric and kinematic considerations, was substantiated as a criterion of the running behaviour of the instantaneous oscillation pattern in the beam cross section, appears to be proportional to the local value of the OFD. Perhaps, it would be too resolute to state that the OFD (and, consequently, the orbital momentum) of a paraxial beam *originates* from the directed motion of the instant field pattern but their interrelation is obvious. In fact, the transverse OFD can be considered as a natural kinematic measure of the "running" features in the instantaneous field motion.

The facts described in this note disclose remarkable aspects of the internal energy flows, their physical nature and underlying mechanisms that seem to escape the attention so far. In contrast, the spin part of the transverse flow density [3] cannot be associated with any spatial motion of the instantaneous field pattern: this property is an additional feature to distinguish the two constituents of the internal flow.

**Acknowledgement**

The author thanks Igor Marienko for helpful assistance.